\documentclass[
pra,
twocolumn,
preprintnumbers,
amsmath,amssymb,
superscriptaddress,
letterpaper,
longbibliography
]{revtex4-2}

\usepackage{color}
\usepackage{graphicx}
\usepackage[utf8]{inputenc}
\usepackage{array} 
\usepackage{amsxtra}
\usepackage{amstext}
\usepackage{amsthm} 
\usepackage{latexsym}
\usepackage{IEEEtrantools} 
\usepackage{verbatim} 
\usepackage{bm} 
\usepackage{hyperref} 
\usepackage{xfrac} 
\usepackage{siunitx}
\usepackage{upgreek}
\usepackage{enumitem}

\begin{document}
\title{Skyrmions in scalar fields of non-Hermitian optical microcavities: \\ spontaneous formation, nonlinear control, and optical forces}

\author{Jan Wingenbach}
\affiliation{Department of Physics and Center for Optoelectronics and Photonics Paderborn (CeOPP), Paderborn University, 33098 Paderborn, Germany}
\affiliation{Institute for Photonic Quantum Systems (PhoQS),Paderborn University, 33098 Paderborn, Germany}

\author{Roman Lebs}
\affiliation{Department of Physics and Center for Optoelectronics and Photonics Paderborn (CeOPP), Paderborn University, 33098 Paderborn, Germany}

\author{Xuekai Ma}
\affiliation{Department of Physics and Center for Optoelectronics and Photonics Paderborn (CeOPP), Paderborn University, 33098 Paderborn, Germany}

\author{Ewan M. Wright}
\affiliation{Wyant College of Optical Sciences, University of Arizona, Tucson, AZ 85721}
\affiliation{Department of Physics, University of Arizona, Tucson, AZ 85721}

\author{Nai H. Kwong}
\affiliation{Wyant College of Optical Sciences, University of Arizona, Tucson, AZ 85721}

\author{Rolf Binder}
\affiliation{Wyant College of Optical Sciences, University of Arizona, Tucson, AZ 85721}
\affiliation{Department of Physics, University of Arizona, Tucson, AZ 85721}

\author{Harald Giessen}
\affiliation{Wyant College of Optical Sciences, University of Arizona, Tucson, AZ 85721}
\affiliation{Physikalisches Institut, Research Center SCoPE, and Integrated Quantum Science and Technology Center (IQST), University of Stuttgart, 70569 Stuttgart, Germany}

\author{Stefan Schumacher}
\affiliation{Department of Physics and Center for Optoelectronics and Photonics Paderborn (CeOPP), Paderborn University, 33098 Paderborn, Germany}
\affiliation{Institute for Photonic Quantum Systems (PhoQS),Paderborn University, 33098 Paderborn, Germany}
\affiliation{Wyant College of Optical Sciences, University of Arizona, Tucson, AZ 85721}

\date{\today}

\begin{abstract}
    Topological textures of light offer powerful routes for structuring optical fields, controlling wave transport, and manipulating matter. Skyrmions, long studied as topological solitons in vector fields, have recently been extended to scalar wave systems, including acoustics, hydrodynamics, and plasmonics. However, their realization in two-dimensional scalar wave propagation with nonlinearities and in quantum fluids remains uncharted. Here, we establish such a Skyrmion framework for scalar fields in optical microcavities. With focus on exciton-polaritons, we show that nonresonant excitation without imposed phase can spontaneously generate isolated Skyrmions and self-organized Skyrmion lattices in a polariton condensate. We trace this mechanism to gain- and loss-induced phase curvature together with outward polariton flow. We further demonstrate that polariton nonlinearities provide all-optical control of these topological textures, enabling switching of the Skyrmion number and reconfiguration of Skyrmion moiré lattices through resonant and nonresonant excitation schemes. These results establish nonlinear non-Hermitian resonators as a versatile platform for the spontaneous generation and active control of scalar topological light fields.
    
\end{abstract}

\maketitle

{\bf Introduction --} Topological textures in continuous fields play a central role across condensed matter, photonics, and fluid dynamics. Among them, Skyrmions, topological excitations and textures characterized by a swirling, particle-like field configuration, have been extensively studied in two-dimensional vector fields, ranging from magnetism~\cite{tokura2020magnetic} and optics~\cite{Du2019,doi:10.1126/science.aau0227, Shen2024, lpor.202501427, Krol:21} to Bose-Einstein condensates (BECs)~\cite{PhysRevLett.81.742}, phonons~\cite{cao2023observation}, and spinor exciton-polariton condensates~\cite{CHENG2024129600}. In this context, exciton-polaritons in two-dimensional planar microcavities as realized in traditional semiconductor platforms, transition metal dichalcogenides (TMDs), perovskites, and organic materials, provide an exceptional optically accessible platform to investigate topological excitations and their dynamics, including targeted nonlinear control~\cite{PhysRevB.88.041308,Ballarini2013,Ma2020NatComm,cookson2021geometric,PhysRevResearch.6.013148, Ma_Nature_Reviews_Physics_2026}. 

Beyond vector fields, recent work has shown that scalar fields can also host Skyrmion textures when embedded into an auxiliary vector representation constructed from amplitude and phase gradients~\cite{smirnova2024water}. Such Skyrmions have been observed in plasmonic nanostructures~\cite{davis2020ultrafast,schwab2024plasmonic, schwab2025skyrmion}, hydrodynamic surface waves~\cite{wang2025topological,che2026twisted}, and surface phonon polariton systems~\cite{schwab2026tunable,mangold2026phonon,Bau2026} where carefully engineered multi-beam interference imposes the required phase structure. These studies raise a fundamental question: can nontrivial Skyrmion topology also emerge spontaneously in scalar complex-valued fields, and can those topologies be controlled by nonlinear interactions rather than external geometrical constraints?

In the present work we answer this question. We develop a general theoretical framework for Skyrmion formation in scalar driven–dissipative microcavity fields, and quantum fluids, such as polariton or Bose-Einstein condensates. As a main outcome of the present study, we show that in a nonresonantly driven exciton–polariton condensate spontaneous formation of Skyrmion topologies can indeed be observed, in a well controlled and tailorable manner. This elevates Skyrmions in scalar fields from externally engineered interference patterns to spontaneously forming and optically reconfigurable topological fields in an active nonlinear photonic platform. This observation opens the path for future explorations into the controlled realization and functionalization of a Skyrmion quantum-fluidic matter, enabling a novel type of photonic Skyrmionics with potential applications in information processing.

While the nonlinear interactions in our system are not required for the existence and formation of such Skyrmion topologies they provide us with a powerful control handle \cite{Ma2020NatComm,PhysRevLett.118.157401}. This we use to demonstrate their efficient nonlinear optical control, including resonant and nonresonant optical excitation schemes. We envision that some of the nonlinear control concepts we introduce may in the future also be transferrable to other platforms such as plasmonic nanostructures combined with an active nonlinear medium. One further avenue to explore may also derive from the optical force fields generated by the Skyrmion textures. These could potentially be used to trap and control atoms or quasi-particles that reside in an adjacent layer. 

\begin{figure}[tb]
    \includegraphics[width=0.5 \textwidth]{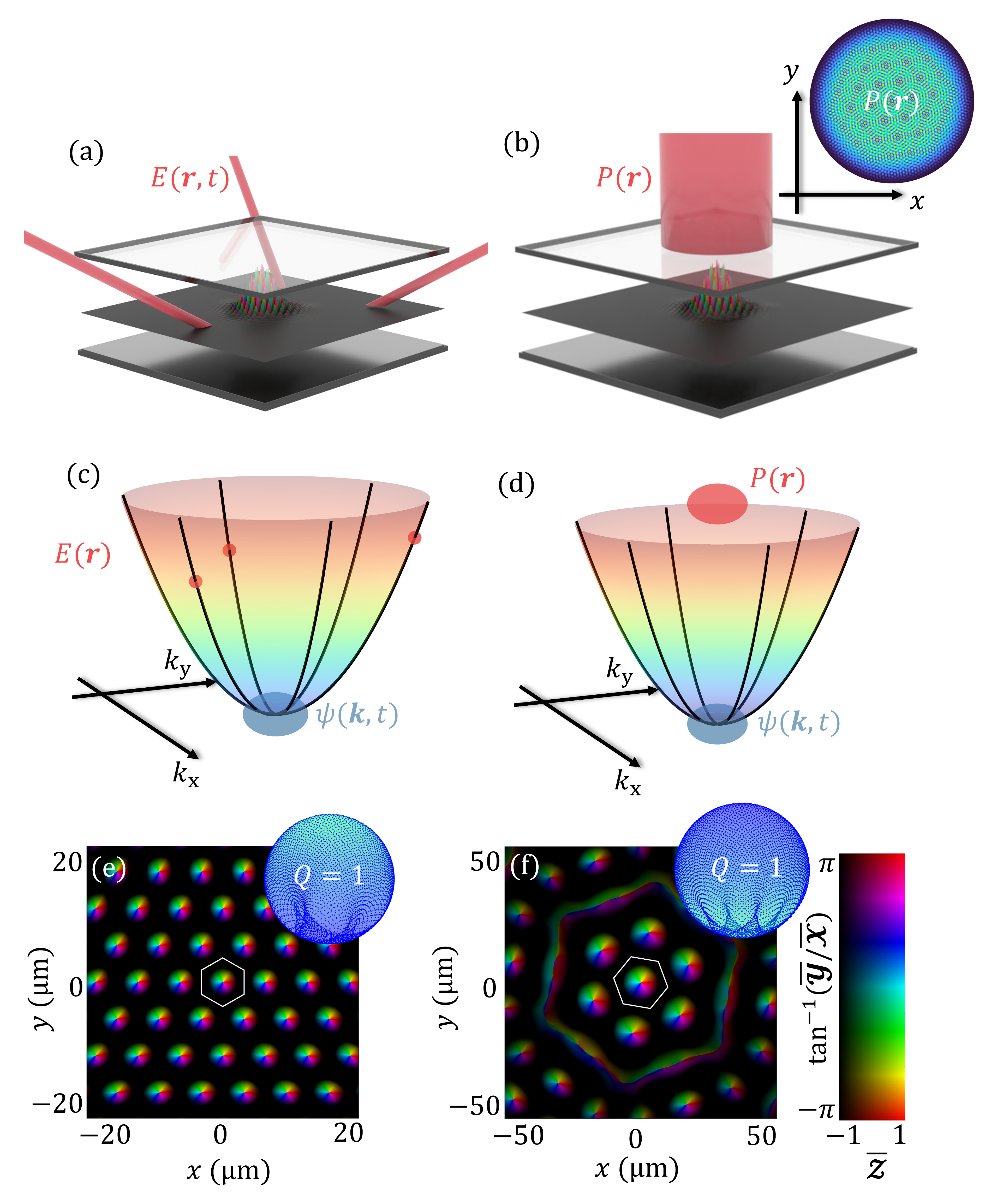}
    \caption{\textbf{Excitation schemes and Skyrmion formation.} Resonant (left column) and nonresonant (right column) excitation of planar microcavity. (Left) $N$ resonant pump beams (a) arranged on a ring in momentum space (c) excite propagating polariton waves that interfere in real space leading to formation of a Skyrmion lattice with corresponding displacement field $\mathcal{R}$ (e); example of hexagonal Skyrmion lattice for three input beams ($N=3$) is shown. (Right) Spatially broad and structured nonresonant pump profile $P(\mathbf{r})$ (b) induces condensate formation in $\psi(\mathbf{k},t)$ near the bottom of polariton dispersion (d). (f) Skyrmion bag in coherent condensed polariton system, spontaneously formed for reservoir with moiré lattice structure as indicated in (b) (for details see text). (e,f) Brightness encodes out-of-plane component $\bar{\mathcal{Z}}$, color encodes in-plane angle $\tan^{-1}(\bar{\mathcal{Y}}/\bar{\mathcal{X}})$ of displacement field $\bar{\mathcal{R}}$. Mapping the normalized field $\bar{\mathcal{R}}$ inside the highlighted areas to the unit sphere provides a Skyrmion with Skyrmion number of $Q= 1$.}
    \label{figure1}
\end{figure}

{\bf Theoretical Foundations --} We model the coherent spatio-temporal dynamics of the polariton condensate based on the extended Gross–Pitaevskii equation coupled to an incoherent excitation reservoir. The dynamics of the condensate order parameter or wave function $\psi = \psi(\mathbf r,t)$ and the incoherent reservoir density $n = n(\mathbf r,t)$ are governed by~\cite{PhysRevLett.99.140402}
\begin{equation}
i\hbar\partial_t\psi= \biggl(-\frac{\hbar^2}{2m}\nabla^2+g_\mathrm{c}|\psi|^2+g_\mathrm{r} n + \frac{i\hbar}{2} [R_\mathrm{c}n-\gamma_\mathrm{c}] \biggr)\psi+E\,,\nonumber
\end{equation}
\begin{equation}
\partial_t n = P - (\gamma_\mathrm{r}+R_\mathrm{c}|\psi|^2)n\,,
\end{equation}
with $\mathbf{r}=(x,y)$. Here, $m = 10^{-4}m_\mathrm{e}$ denotes the effective particle mass of the lower polariton branch. The interaction strength between condensed polaritons is $g_\mathrm{c} = 0.6~\mathrm{\upmu eV\upmu m^2}$, while $g_\mathrm{r} = 2g_\mathrm{c}$ characterizes the condensate-reservoir interaction. Coherent (resonant) driving of the condensate enters via the complex-valued field $E = E(\mathbf{r},t)$, while incoherent (nonresonant) pumping is described by the real-valued field $P = P(\mathbf r,t)$, which drives the reservoir. Stimulated in-scattering with $R_\mathrm{c} = 0.01~\mathrm{ps^{-1}\upmu m^2}$ feeds the condensate from the reservoir which in turn depletes the reservoir density. Dissipation enters through the condensate decay $\gamma_\mathrm{c} = 0.3~\mathrm{ps^{-1}}$, and reservoir decay $\gamma_\mathrm{r} = 1.5\gamma_\mathrm{c}$. Here, these parameters are chosen to be in a realistic range for typical GaAs microcavities~\cite{PhysRevB.100.035306,PhysRevLett.118.016602}. Moreover, the results presented below remain qualitatively the same under variations of the system parameters. We also tested that our results are robust against random disorder potentials at least up to $0.2~\mathrm{meV}$ with correlation lengths of about $1~\mathrm{\upmu m}$.

To calculate the Skyrmion structure for stationary states $\psi_\mathrm{s}$ (for persistent coherent or incoherent driving) with frequency $\omega$ in the rotating frame, we adapt and apply the mathematical approach given in Ref.~\cite{smirnova2024water}. To this end we define the complex-valued displacement field $R=(\partial_x,\partial_y,1)^\mathrm{T}\psi_\mathrm{s}$. This construction embeds the complex scalar field into a three-component vector field whose orientation encodes local amplitude and phase gradients. From this, the real-valued displacement field $\mathcal{R}=(\mathcal{X},\mathcal{Y},\mathcal{Z})^\mathrm{T} = \Re(\mathrm{e}^{i\omega t}R)$ is obtained, where $\mathrm{e}^{i\omega t}$ eliminates the temporal evolution of $\psi_\mathrm{s}$. This elimination permits calculation of a stationary topological charge $Q$, which is a crucial requirement for example for the Skyrmion switching investigated below. The topological charge of the Skyrmion ("Skyrmion number") is calculated over one Skyrmion cell by
\begin{equation}
\label{eq:skyrmion}
Q=\frac{1}{4\pi}\iint\bar{\mathcal{R}} (\partial_x\bar{\mathcal{R}}\times\partial_y\bar{\mathcal{R}})\mathrm{d}x\mathrm{d}y,    
\end{equation}
with $\bar{\mathcal{R}}=\mathcal{R}/|\mathcal{R}|$. The bounds of the integration domain, i.e. the Skyrmion cell, are defined by critical points in the displacement field in agreement with previous Skyrmion-related works~\cite{Du2019, He2024,schwab2024plasmonic,Shen2024,schwab2025skyrmion,wang2025topological,che2026twisted,neuhaus2026linking}. Mapping the directions of $\bar{\mathcal{R}}$ to the unit sphere visualizes the Skyrmion vector topology. In similar fashion, the topology of meron (half-Skyrmion) structures can be visualized via the spin density $S = (S_x,S_y,S_z)^\mathrm{T} \propto \Im(R^*\times R)$. While outside the scope of this work, we use this to also validate the existence of meron lattices for three-wave interference in SM Note S3.

It is essential to emphasize that the Skyrmions considered here are not spin-texture Skyrmions associated with internal polarization degrees of freedom, as they are widely investigated in many systems including spinor-polariton condensates~\cite{PhysRevLett.110.016404, donati2016twist, PhysRevB.94.045315, PhysRevE.104.054216, CHENG2024129600}. Instead, they arise from a mapping of a complex-valued scalar field into a three-component vector field constructed from spatial derivatives. This defines a continuous mapping $\mathbb{R}^2 \to S^2$, allowing a topological charge analogous in mathematical structure to that of conventional Skyrmions in vector fields, when evaluated over an appropriately chosen Skyrmion cell or integration domain. While the vector field in the present work is derived from amplitude and phase gradients rather than intrinsic spin, the resulting topology is genuine in the sense that it is invariant under smooth deformations of the scalar field that preserve boundary conditions. This construction therefore represents a scalar-field realization of local Skyrmion topology~\cite{smirnova2024water}. As in other optical and interference-based Skyrmion textures, the integer-valued charge is meaningful when the integration domain is chosen to match the physically defined cell of the texture; arbitrary deformations of this domain generally lead to non-integer values but are not to be interpreted as invalidation of the local topological structure.

Moreover, the complex-valued displacement field $R$ can be related to the phase- and intensity-gradient forces of the polariton field. In SM Note S1 the relation between these forces and the in-plane displacement field $\mathcal{R}_2=(\mathcal{X},\mathcal{Y})$ is derived. The phase-gradient force is proportional to
\begin{equation}
    F_\phi \propto I(x,y)\, \nabla \phi(x,y)
    = \Im \left\{ \psi_s^*(x,y)\, R_2(x,y) \right\}
    \label{eq:force_phi}
\end{equation}
and the intensity-gradient force is proportional to
\begin{equation}
    F_I \propto \nabla I(x,y)
    = 2 \Re \left\{ \psi_s^*(x,y)\, R_2(x,y) \right\}\,.
    \label{eq:force_i}
\end{equation}
Accurate post-processing and calculation of the required gradient fields relies on the accurate evaluation of spatial gradients along both $x$ and $y$, requiring sufficiently finely discretized grids. At the same time, the simulation domain must be sufficiently large to accommodate the entire pumping region while preventing boundary artifacts. For persistent driving (resonant coherent or off-resonant incoherent, depending on the scenario investigated as detailed below), here we evolve the system in time until stationary behavior is observed, on large two-dimensional grids with on the order of $\sim 10^7$ to $\sim 10^8$ grid points. To render these computationally intense calculations feasible, we use PHOENIX, our high-performance solver for two-dimensional nonlinear Schrödinger equations~\cite{wingenbach2025phoenix}.

\begin{figure}[tb]
    \includegraphics[width=0.5 \textwidth]{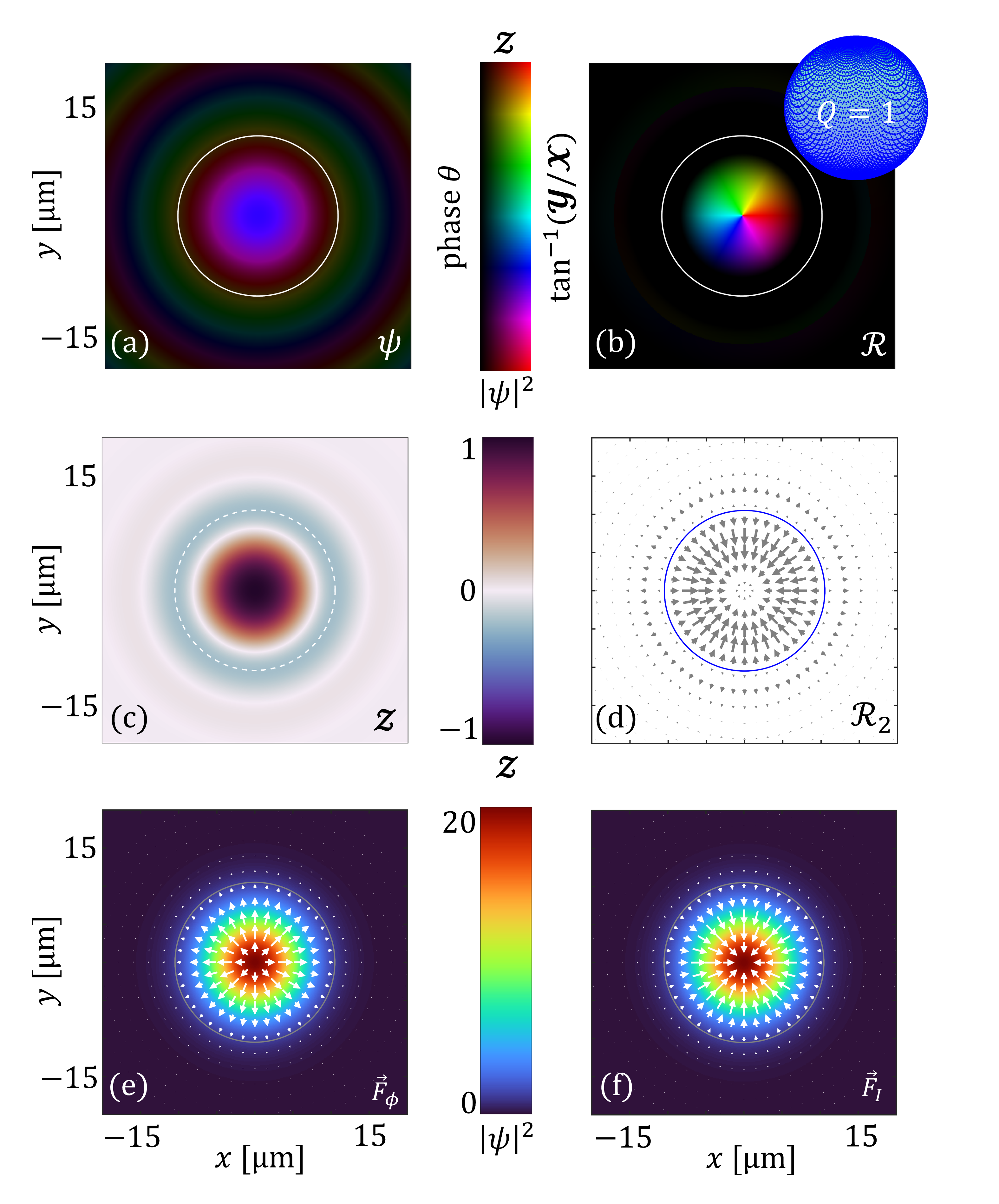}
    \caption{\textbf{Spontaneous Skyrmion formation for localized nonresonant pump.} (a) An individual continuous Gaussian-shaped (in real space) nonresonant pump beam $P(\mathbf{r})$ induces a stationary condensate; density (brightness), phase (color). Finite lifetime and reservoir-induced effective external potential induce radial outflow and phase gradients. (b) Corresponding displacement field $\mathcal{R}=(\mathcal{X},\mathcal{Y},\mathcal{Z})^T$ for stationary condensate wavefunction, normalized to peak density. (c) Out-of-plane component $\mathcal{Z}$ showing sign reversal. (d) In-plane component $\mathcal{R}_2=(\mathcal{X},\mathcal{Y})$ revealing angular winding. Mapping $\bar{\mathcal{R}}$ to the unit sphere yields Skyrmion charge $Q = 1$ (radius marked in each panel). (e) Phase-gradient force $\vec{F}_\phi$ and (f) density-gradient force $\vec{F}_I$ (vector fields) and density distribution (color) in the Skyrmion vicinity. $\vec{F}_I$ acts inward towards the Skyrmion center and $\vec{F}_\phi$ away from the center. Note that a small brightness offset is used in (a) to visualize the radial phase gradient around the condensate induced by non-Hermiticity.}
    \label{figure2}
\end{figure}

{\bf Skyrmions from Linear Interference --} First we reproduce established Skyrmion and meron patterns arising from three-wave interference and more complex moiré-type interference configurations as previously reported in hydrodynamic~\cite{wang2025topological,che2026twisted} and plasmonic wave systems~\cite{schwab2024plasmonic, schwab2025skyrmion}. In planar optical microcavities resonant driving can be used to generate the desired propagating waves. The source inducing resonant excitation, $E(\mathbf r,t)$, then consists of $N$ Gaussian-profiled (in real space) excitation beams arranged on a ring in real- and reciprocal-space as sketched in Fig.~\ref{figure1}(a); mathematical description given in SM Note S2. Three-wave interference is obtained with three spots separated by $\ang{120}$ angles~\cite{wang2025topological}. A moiré lattice as in Ref.~\cite{wang2025topological} is realized with a total of $N=18$ excitation beams forming three hexagons with relative angle of $\phi_{12} = \phi_{23} = \ang{9.4}$ (as indicated in Fig.~4(a) below). 

In Fig.~\ref{figure1}(e) the interference induced real-valued displacement field in the center of the spatial domain is shown. Here, brightness encodes the $\bar{\mathcal{Z}}$ component and coloring encodes $\mathrm{atan^{-1}}(\bar{\mathcal{Y}}/\bar{\mathcal{X}})$. The bright features that contain all the colors represent Skyrmion signatures with Skyrmion number $Q=1$ (the hexagonal unit cell used for Skyrmion number evaluation is highlighted in white). In agreement with Refs.~\cite{smirnova2024water,schwab2024plasmonic,wang2025topological,che2026twisted} the unit-cell boundary is defined by the points in radial direction (away from the respective Skyrmion center) where the displacement field changes sign. For illustration, in Fig.~\ref{figure1}(f) the displacement field for a moiré lattice excitation is illustrated (albeit here obtained for a different, off-resonant excitation setup as discussed in detail in the following section). For a detailed analysis of the corresponding wavefunction, in-plane displacement field and meron structure are shown in SM Note S4. The Skyrmion structures observed here are equivalent in topology and displacement field to those reported for plasmonic or water wave systems~\cite{schwab2024plasmonic,wang2025topological,che2026twisted}.

{\bf Spontaneous Formation of Skyrmions --} As demonstrated in the previous section, Skyrmion patterns in wave systems are typically engineered through carefully designed beam or wave interference, with input wave geometries that impose specific phase relations between multiple propagating coherent waves or beams. 

As an essential and hitherto unexpected result of the present study, here we demonstrate that such external phase control is not necessarily required when turning to non-Hermitian exciton–polariton condensates, where condensate formation from noise and phase synchronization over large spatial domains may occur spontaneously. For that system, only using nonresonant excitation that does not carry phase information, we demonstrate that Skyrmions can emerge spontaneously in the driven–dissipative dynamics of the system. As a side note we would like to mention that this nonresonant excitation method is in principle also compatible with spatially structured electrical (rather than optical) pumping of the condensate.

Let us first consider a single Gaussian (in real space) nonresonant pump, with no imposed phase information. The pump feeds the incoherent reservoir $n(\mathbf{r},t)$, which (if above condensation threshold) in turn populates the condensate via stimulated scattering [cf. Fig.~\ref{figure2}(a)] triggered by initial random noise. Once reaching a stationary state, condensed particles generated inside the localized pumping region propagate radially outward before decaying. This outward flux leads to a non-vanishing probability current $\mathbf{j} = (\hbar/m)\,\mathrm{Im}(\psi^* \nabla \psi)$ and therefore a radial phase gradient $\nabla \phi \propto \mathbf{j}/|\psi|^2$. The resulting steady-state solution for the coherent condensate exhibits oscillatory density modulations (in radial direction) and continuously varying phase around the pump center. The associated displacement field $\mathcal{R}$ in Fig.~\ref{figure2}(b) develops sign changes in its out-of-plane component $\mathcal{Z}$ in Fig.~\ref{figure2}(c) and a Skyrmion-cell border in its in-plane component $\mathcal{R}_2=(\mathcal{X},\mathcal{Y})$ in Fig.~\ref{figure2}(d). As a result, the normalized vector field shows the characteristic vector topology to span the full unit sphere and form a Skyrmion with $Q=1$. The Skyrmion-cell boundary is defined in agreement with optical Skyrmions observed in the photonic spin structure~\cite{Du2019, He2024, Shen2024,neuhaus2026linking}. In agreement with Ref.~\cite{neuhaus2026linking}, increasing the integration radius beyond the inner Skyrmion cell in evaluating Eq.~\eqref{eq:skyrmion}, the outcome alternates between $0$ and $1$, as plotted in Fig.~S6 of the SM Note 7. A comprehensive approach to  topological invariants for optical Skyrmions has recently been provided in Ref.~\cite{neuhaus2026linking}.

In addition to the crucial observation of spontaneously formed Skyrmion textures in the displacement field, the resulting optical forces~\eqref{eq:force_phi} and~\eqref{eq:force_i} can be evaluated. These are displayed as vector fields in front of the density distribution in Fig.~\ref{figure2}(e,f). While the phase-gradient force acts outward from the Skyrmion center, the density-gradient force acts inward towards the Skyrmion center. This observation may open up the possibility to trap atoms or quasi-particles (that may reside in an adjacent layer) with or inside the Skyrmion texture. Combining the possible quantum nature of the trapped particles with the topological character of the Skyrmion, with sufficiently strong coupling, this could motivate future studies into novel quasi-particles with intriguing fractional anyonic statistics~\cite{wright1994optical, PhysRevB.103.035103,preece2025harnessing, PhysRevLett.101.260501, PhysRevLett.117.205303}. It is worth noting that the phase-gradient force (albeit pointing away from the Skyrmion center for an isolated Skyrmion) may be used to trap particles in between Skyrmions in extended lattice structures (illustrated for a bag of Skyrmions in a moiré lattice in SM Note S5). We further note that the absolute and relative strength of phase-gradient and density-gradient forces will depend on the specific physical realization and material platform used. In this context also other nonlinear topological modes, such as vortex states with their winding phase-gradient force fields, may be worth exploring in the future.

The system dynamics described by the driven–dissipative Gross–Pitaevskii equations is intrinsically nonlinear due to condensate–reservoir coupling and intra-condensate interaction. However, it is important to note that the emergence of the Skyrmion topology does not rely on these nonlinearities. We have verified that nontrivial Skyrmion charge persists in the limit $g_\mathrm{c}=g_\mathrm{r}=0$, as well as in the low-density regime where gain saturation is negligible ($R_\mathrm{c}|\psi|^2 \ll \gamma_\mathrm{r}$). Thus, while the dynamical equations remain nonlinear in general, the mechanism underlying spontaneous Skyrmion formation is fundamentally rooted in phase gradients generated by localized gain and outwards propagation with finite lifetime. In SM Note 6 we demonstrate that a non-Hermitian Schrödinger equation provides the analytical solution close to the condensation threshold that relates the Skyrmion properties of the solution to the phase curvature induced by the non-Hermiticity. This is validated in the full numerical calculations as the Skyrmion size is anti-proportional to $m$ and $\gamma_\mathrm{c}$, which directly control the radial propagation distance of the polaritons. On the other hand nonlinear interactions, which are a particular strength of our microcavity polariton platform, provide powerful knobs to control the Skyrmions formed as we will demonstrate further below.

Another important aspect to note is that for resonant excitation with a Gaussian-shaped beam (in real space) centered around $\mathbf{k}=0$ Skyrmion textures are not found. In the resonantly driven case, the external field fixes the condensate phase, suppressing self-organization of the required phase gradients. The resulting steady state that is reached for cw coherent excitation in normal incidence inherits the Gaussian envelope of the pump, and the out-of-plane displacement component remains strictly positive, in turn yielding a trivial topology.

\begin{figure}[t]
    \includegraphics[width=0.5 \textwidth]{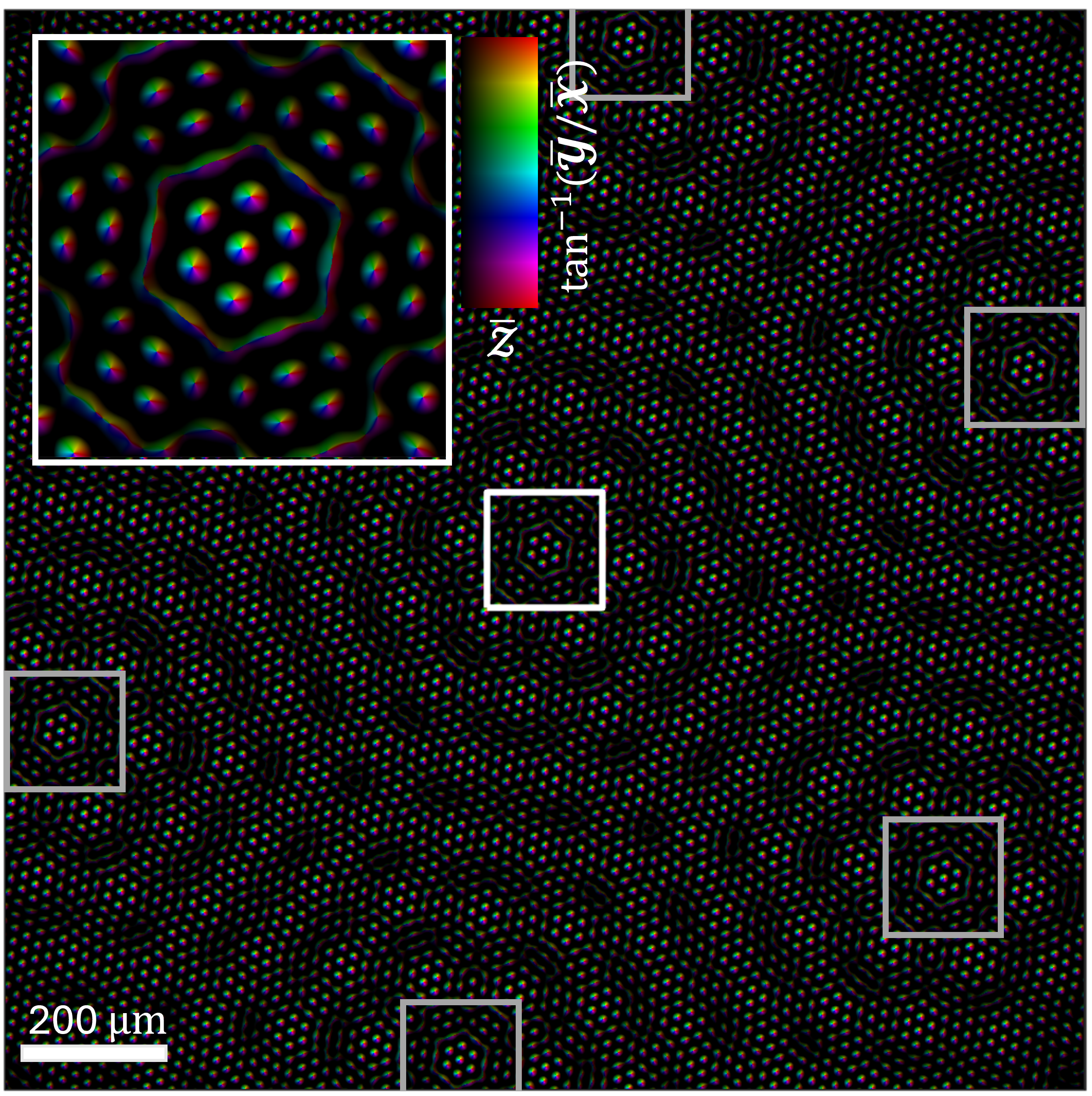}
    \caption{\textbf{Spontaneously formed Skyrmion bags in polariton condensate in a moiré-shaped reservoir.} On the large two-dimensional spatial grid with $\sim10^8$ grid points multiple Skyrmion bags are resolved as indicated by grey squares. The Skyrmion structure is shown via the displacement field $\mathcal{R}=(\mathcal{X},\mathcal{Y},\mathcal{Z})^T$($\mathcal{Z}$ encoded in brightness and $\mathrm{atan^{-1}}(\mathcal{Y}/\mathcal{X})$ in color). The inset shows the central Skyrmion bag. Details of structured excitation reservoir in text.}
    \label{figure3}
\end{figure}

Based on the observed spontaneous formation of a Skyrmion for a single Gaussian-shaped nonresonant pump we now explore more complex spatial profiles of the nonresonant pump feeding the condensate. The nonresonant pump profile density is generated by superposition of $\mathrm{cos}$-shaped waves with planar wave fronts with different spatial orientation (details given in SM Note S3). To prevent boundary effects in our finite computational domain, a very broad super-Gaussian envelope is applied to the pump profile. One possible boundary effect is that due to polariton flow at the envelope's edges, the outermost Skyrmions can be distorted, which is why we focus on the observations in the center of the spatial domain. The three-wave-interference results with nonresonant excitation are displayed in detail in SM Note S4. We note that also with nonresonant pumping they very closely resemble the results for resonant coherent excitation as displayed in Fig.~\ref{figure1}(e). 

For the nonresonant excitation we now turn to a more complex spatial profile of the pump for which phase-synchronization of the condensate -- in a way such that spatially extended Skyrmionic textures would form -- is not trivially expected. Here it is important to emphasize that while the nonresonant pump may define an amplitude landscape, the condensate phase, phase gradients, and resulting Skyrmion topology emerge spontaneously in the driven-dissipative dynamics of the condensate from initial noise. To demonstrate this kind of self-organization on large scales and show the periodic nature of the system we impose the moiré grid structure of Fig.~1(f) on a very large spatial grid. The resulting spontaneously formed Skyrmion structure is displayed in Fig.~\ref{figure3} on a $1500\times1500~\mathrm{\upmu m^2}$ grid, showing the periodicity of the moiré structure with repetitions of the central Skyrmion bag also far away from the origin. 

For completeness we note that similar Skyrmionic structures in the condensate can also form in non-optically induced, external potential lattice structure excited with a spatially homogeneous nonresonant pump (not shown). In that case careful parameter optimization is required. We further note that in the nonresonant excitation scenarios Skyrmion formation is more resilient against pumping imperfections. For resonant excitation, wave vector deviations larger than $1\%$ can lead to destruction of the Skyrmion patterns. As no external phase control is required for nonresonant pumping, Skyrmion patterns are preserved for nonresonant pump amplitude deviations of up to $\sim30\%$ and deviations in the $\mathrm{cos}$ argument for lattice generation of more than $5\%$. 

\begin{figure}[tb]
    \includegraphics[width=0.5 \textwidth]{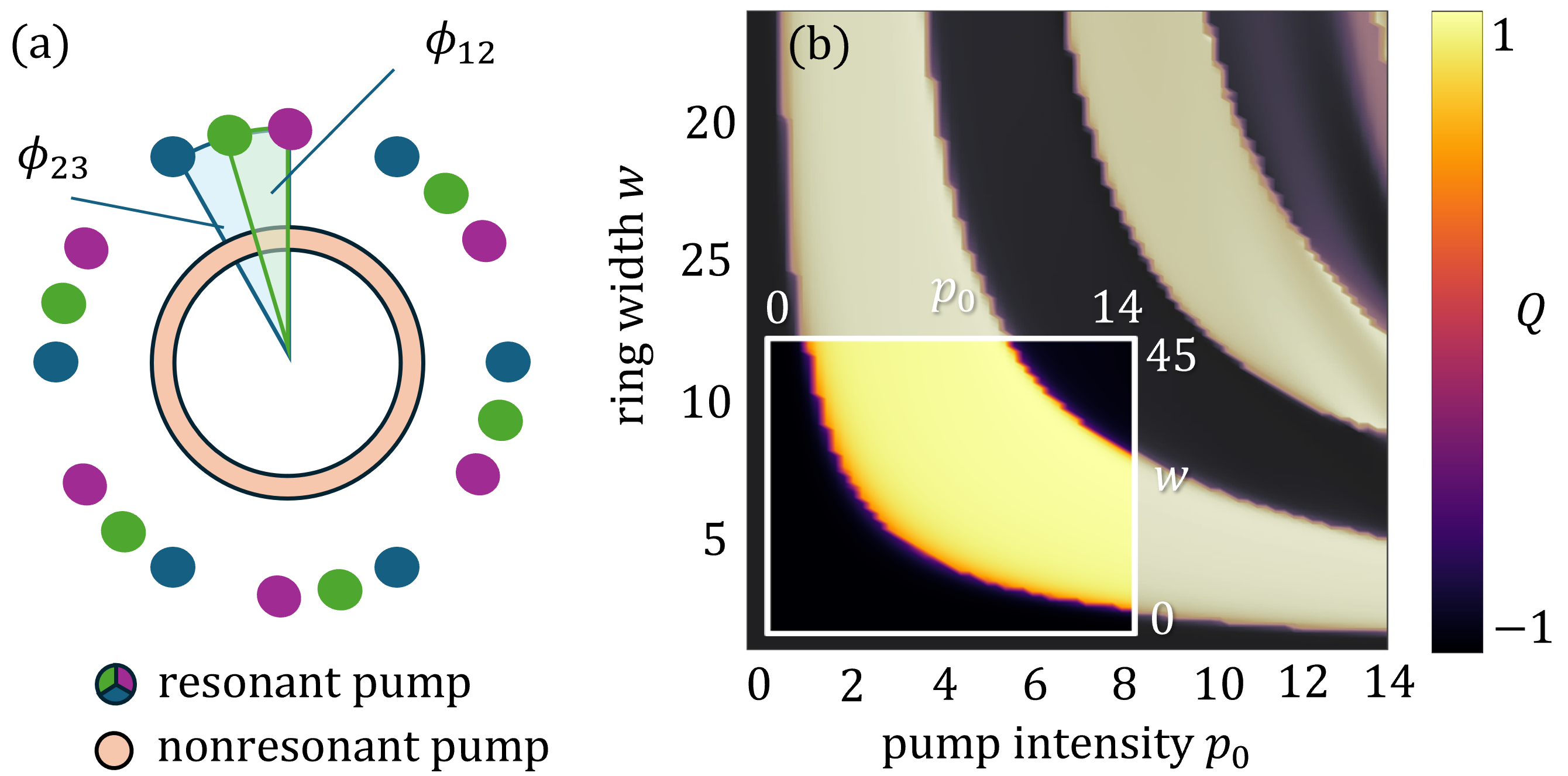}
    \caption{\textbf{Nonresonant optical control of Skyrmion number.} (a) Optical excitation and switching setup: Skyrmion bags driven with resonant beams (dots) forming twisted hexagons (marked by color). A nonresonant ring pump (orange) creates a phase-shift to switch the Skyrmion numbers inside the ring. (b) For a twisting-angle $\phi_{12} = \phi_{23} = \ang{19.4}$ a single-Skyrmion bag is switched by varying ring-pump intensity $p_0$ and width $w$.}
    \label{figure4}
\end{figure}

\begin{figure}[tb]
    \includegraphics[width=0.5 \textwidth]{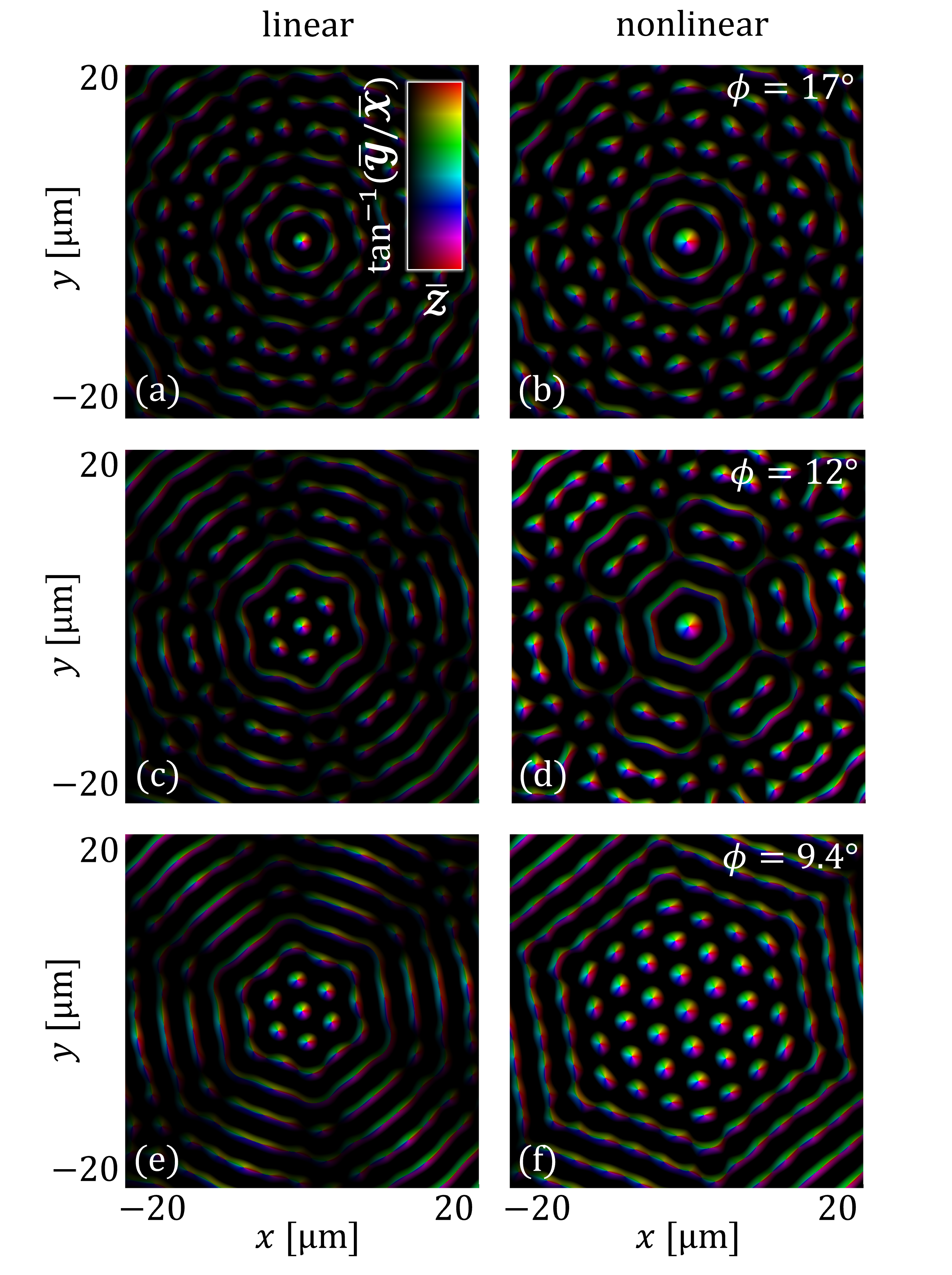}
    \caption{\textbf{Skyrmion manipulation enabled by Kerr nonlinearity}. Left: resonantly excited Skyrmion bags in moiré lattices induced by interference of 18 resonant beams (cf. sketch in Fig.~4(a)) with twist angles of (a) $\phi=\ang{17}$, (c) $\phi=\ang{12}$, and (e) $\phi=\ang{9.4}$ in the linear regime. Right: same as left but in the nonlinear regime with elevated beam intensities. Nonlinearity induces increased size of individual Skyrmions as visible in (b) and significant alterations to bag structures in (d) and (f).}
    \label{figure5}
\end{figure}

{\bf Nonlinear Control of Skyrmions --} While Skyrmions emerge already in the linear wave dynamics of the system, the intrinsic nonlinearities of the polariton condensate provide a powerful control knob over their topology and spatial structure. We now demonstrate two approaches illustrating the possibilities. The first lets us switch the topological Skyrmion charge in a controlled manner (Fig.~\ref{figure4}). The second leads to global reorganization of the Skyrmion patterns in the nonlinear regime (Fig.~\ref{figure5}). We begin with resonant excitation inducing a moiré Skyrmion lattice (also see SM Note S2). For relative angles between hexagons of $\phi_{12} = \phi_{23} = \ang{19.4}$, a central bag containing a single Skyrmion forms in the center as shown in Fig.~\ref{figure5}(a) below.

To actively control the topology of this central Skyrmion we now introduce an additional nonresonant ring-shaped pump surrounding the central region as sketched in Fig.~\ref{figure4}(a). This ring does not directly interfere with the resonantly injected waves. Instead, it modifies their propagation before they meet or overlap at the center due to a reservoir-induced effective external potential. Depending on strength $p_0$ and width $w$ of the reservoir ring, this induces a controllable phase delay for the propagating waves. 

Figure~\ref{figure4}(b) shows the resulting Skyrmion number in the $(p_0,w)$-plane for varying accumulated phase shift. In the bright area the Skyrmion number for $g_\mathrm{c}= 0.6~\mathrm{\upmu eV\upmu m^2}$ (consistent with the parameters used in the present paper) is shown. We display the extension for larger nonlinearity parameter $g_\mathrm{c}=6~\mathrm{\upmu eV\upmu m^2}$ in the grayed-out area. We note that for $g_\mathrm{c}= 0.6~\mathrm{\upmu eV\upmu m^2}$ in the parameter regime belonging to the grayed-out area condensation inside the ring in observed, rendering the system dynamics much more complicated. In the $(p_0,w)$-plane, as the nonlinear phase delay increases, the topological charge of the central Skyrmion switches abruptly. The transitions occur when the accumulated phase shift 
\begin{align}
    \varphi \approx \frac{g_\mathrm{r}}{\hbar v_\mathrm{g}}\int_\mathrm{ring} n(\mathbf{r})\mathrm{d}r
\end{align}
changes the interference condition in the center. Here, $v_\mathrm{g} = \hbar k/m$ is the group velocity of the propagating polariton wave on the parabolic dispersion. These results demonstrate that optically induced changes (by nonresonant pumping) to the propagation outside the interference region enables controlled switching of the topological information contained in the central Skyrmion (without modifying the resonant pumping geometry itself). 

We further note that for sufficiently strong ring-pump, the induced potential forms an optical trap, leading to multistability inside the ring (not shown). This regime where the nonlinear dynamics becomes more complex is beyond the scope of the present work but highlights the broader potential for nonlinear topological engineering in our microcavity polariton system.

Beyond discrete topological switching, Kerr-type nonlinearities directly reshape the Skyrmion lattice itself. For resonant excitation only, increasing the pump intensity enhances the effective intra-condensate interaction, interaction-induced blueshifts, and thereby re-normalizes the local wave vectors participating in the interference. The moiré patterns rely sensitively on precise wave vector control; even moderate nonlinear shifts can significantly alter the resulting Skyrmion topology. Because the interference condition depends on relative wave vectors $k$ and interactions re-normalize the dispersion, effective shifts in $k$ alter the moiré periodicity.

Figure~\ref{figure5} compares several configurations with different relative twist angles in the linear and in the nonlinear regimes (coherent excitation). For $\phi=\ang{17}$, the single-Skyrmion bag expands substantially as the nonlinearity increases [cf. Figs.~\ref{figure5}(a) and (b)], consistent with a reduction of the effective in-plane wave vector magnitude. More strikingly, configurations that appear nearly identical in the linear regime [cf. Figs.~\ref{figure5}(c) and (e)] evolve into qualitatively distinct topological structures at elevated pump intensity [cf. Figs.~\ref{figure5}(d) and (f)].

{\bf Conclusions --} We have presented a theoretical framework for Skyrmion analysis, formation, and nonlinear control in scalar two-dimensional wave systems, with a focus on exciton-polariton condensates in planar semiconductor microcavities. Unlike previous Skyrmion realizations based on imposed interference geometries, here the topology can emerge spontaneously from non-Hermitian condensate dynamics and can be reconfigured through optical nonlinearities, without the need to externally control the specific phase structure and without the need for resonant coherent optical excitation. In particular, we demonstrate controlled switching of topological charge and collective reconfiguration of Skyrmion moiré lattices by introducing additional spatially structured nonresonant optical pumping or by utilizing Kerr-type nonlinearities at elevated pumping intensities. We also show how the displacement fields in which these Skyrmionic topologies are observed can be linked to the physical quantities of intensity-gradient and phase-gradient optical forces.


Our findings combine linear interference-based Skyrmion generation with nonlinear topological engineering, establishing driven-dissipative polariton systems as a versatile and promising platform for studying collective Skyrmion states and their dynamical evolution. Future work may also explore the dynamical stability of spontaneously formed Skyrmions, the role of quantum fluctuations \cite{LudersPRL2023} and condensate lifetime, and the interplay of nonlinear non-Hermitian physics \cite{WingenbachACSPhotonics2026,KwongNatComm2026} and Skyrmion topology. Our study also reveals direct implications for possible experimental realizations in planar photonic resonators, but also in plasmonic systems, and other two-dimensional (quantum) fluids, where the nonlinear control of topological excitations could enable new approaches to information processing and programmable photonics. While going beyond the scope of the present paper we briefly note that we also find that saturable gain may lead to inversion of the Skyrmion structures observed at elevated densities, and multistability may provide additional intriguing control mechanisms for the Skyrmion topologies. Additionally, phase synchronization in specifically designed spontaneously formed Skyrmion lattices may permit controlled switching of Skyrmion phases and complex self-organization by inter-Skyrmion coupling. 

Projecting further ahead, the introduced relation between force fields and Skyrmions may open up new possibilities to trap and manipulate quantum particles inside the Skyrmion force field with mutual coupling, which may motivate future studies into (lattices of) anyons with mixed statistics as topologically protected entities. Along the same lines also the optical force fields generated by other topological nonlinear excitations, such as vortices~\cite{ma2018multi,Ma_Nature_Reviews_Physics_2026, zhai2024splitting} with their winding phase-gradient forces, should be explored.

{\bf Acknowledgements --} The authors gratefully acknowledge financial support for the Paderborn groups, by the Deutsche Forschungsgemeinschaft (DFG, German Research Foundation) through Grant No.~519608013 and the transregional collaborative research center TRR 142 (Projects A04, Grant No. 231447078), and for the Arizona group from the US National Science Foundation (NSF) under Grant No. DMR-1839570. The authors gratefully acknowledge the computing time provided to them on the high-performance computers at the NHR Center PC$^2$ under project hpc-prf-pdsm.

%

\clearpage

\onecolumngrid

\setcounter{equation}{0}
\setcounter{figure}{0}
\setcounter{section}{0}

\renewcommand{\thefigure}{S\arabic{figure}}
\renewcommand{\theequation}{S\arabic{equation}}
\renewcommand{\thesection}{S\arabic{section}}

\begin{center}
    {\Large \bfseries Supporting Information:}\\[1em]
    {\Large Skyrmions in scalar fields of non-Hermitian optical microcavities: \\ spontaneous formation, nonlinear control, and optical forces}
\end{center}

\section{Light forces and displacement field relations}

The displacement field of the stationary-mode polariton field $\psi_s$ is defined as
\begin{equation}
    R = \begin{pmatrix} \partial_x \psi_s \\
    \partial_y \psi_s
    \\ \psi_s
    \end{pmatrix}
\end{equation}
and the polariton field is defined by
\begin{equation}
    \psi_s(x,y) = A(x,y) \mathrm{e}^{i\phi(x,y)}\, ,
\end{equation}
with an amplitude and phase distribution $A(x,y)$ and $\phi(x,y)$, as well as the polariton density $I(x,y) = |\psi_s(x,y)|^2 = A^2(x,y)$. Insertion into the displacement field provides
\begin{equation}
    R= e^{i\phi}
    \begin{pmatrix}
        \partial_x A + i A \partial_x \phi \\
        \partial_y A + i A \partial_y \phi \\
            A
    \end{pmatrix}
\end{equation}

and respectively

\begin{equation}
    \psi_s^* R = A
    \begin{pmatrix}
        \partial_x A + i A \partial_x \phi \\
        \partial_y A + i A \partial_y \phi \\
        A
    \end{pmatrix}\,.
\end{equation}

The in-plane displacement field $R_2$, i.e. the ($x$,$y$) components, then reads
\begin{equation}
    \psi_s^* R_2 = A
    \begin{pmatrix}
        \partial_x A + i A \partial_x \phi \\
        \partial_y A + i A \partial_y \phi 
    \end{pmatrix}\,.
\end{equation}

Hence, we can write down the relation between the displacement field and the light forces:

The phase-gradient force is proportional to
\begin{equation}
    F_\phi \propto I(x,y)\, \nabla \phi(x,y)
    = \Im \left\{ \psi_s^*(x,y)\, R_2(x,y) \right\}
\end{equation}
and the intensity-gradient force is proportional to
\begin{equation}
    F_I \propto \nabla I(x,y)
    = 2 \Re \left\{ \psi_s^*(x,y)\, R_2(x,y) \right\}\,.
\end{equation}

\section{Resonant pump profile definition}
In this section we provide the mathematical description of the resonant pump profile used to investigate both three-wave interference and moiré lattices. The resonant pump profile is described by
\begin{align}
E(\mathbf r,t)
= E_0\,e^{i\omega f(t)}
\sum_{j=1}^{N}
\exp\!\left[
-\frac{|\mathbf r-\mathbf r_j|^2}{2\sigma^2}
+ i\,\mathbf k_j\!\cdot\!\mathbf r
\right],
\label{eq:resonant}
\end{align}

where
\[
\mathbf r=(x,y), \qquad
\mathbf r_j=(x_0(j),y_0(j)), \qquad
\mathbf k_j=(k_x(j),k_y(j)).
\]

$E_0$, $\sigma$, and $\omega=\frac{\hbar|\mathbf k|^2}{2m}$ are the amplitude, width, and frequency of each resonant pump spot. $f(t)$ holds the temporal envelope of the pump. $\mathbf r_j$ and $\mathbf k_j$ hold the positions of each pump spot in direct and reciprocal space.

\section{Non-resonant pump profile definition}
In this section we provide the mathematical description of the nonresonant pump profile used to investigate spontaneous Skyrmion formation in both three-wave interference and moiré lattices in polariton systems. The nonresonant pump profile is described by
\begin{equation}
    P(x,y) = \sum^M_{m=0}\sum^N_{n=0}0.5[1+\mathrm{cos}(k(x\mathrm{cos}(\alpha_{n,m}))+y\mathrm{sin}(\alpha_{n,m}))],
\end{equation}
with $\alpha_{n,m} = \theta_n+m\phi$ and
\[\theta_n = \frac{2\pi n}{3} \qquad
M = 0, \qquad
N = 5.\]
for the three-wave interference setting and
\[\theta_n = \frac{n\pi}{3}, \qquad
M = 2, \qquad
N = 5, \qquad
\phi = \ang{9.4}.
\]
for the moiré lattice setting.

\section{Details about Skyrmion patterns in three-wave interference and moiré lattices}

\begin{figure}[tb]
    \includegraphics[width=0.5 \textwidth]{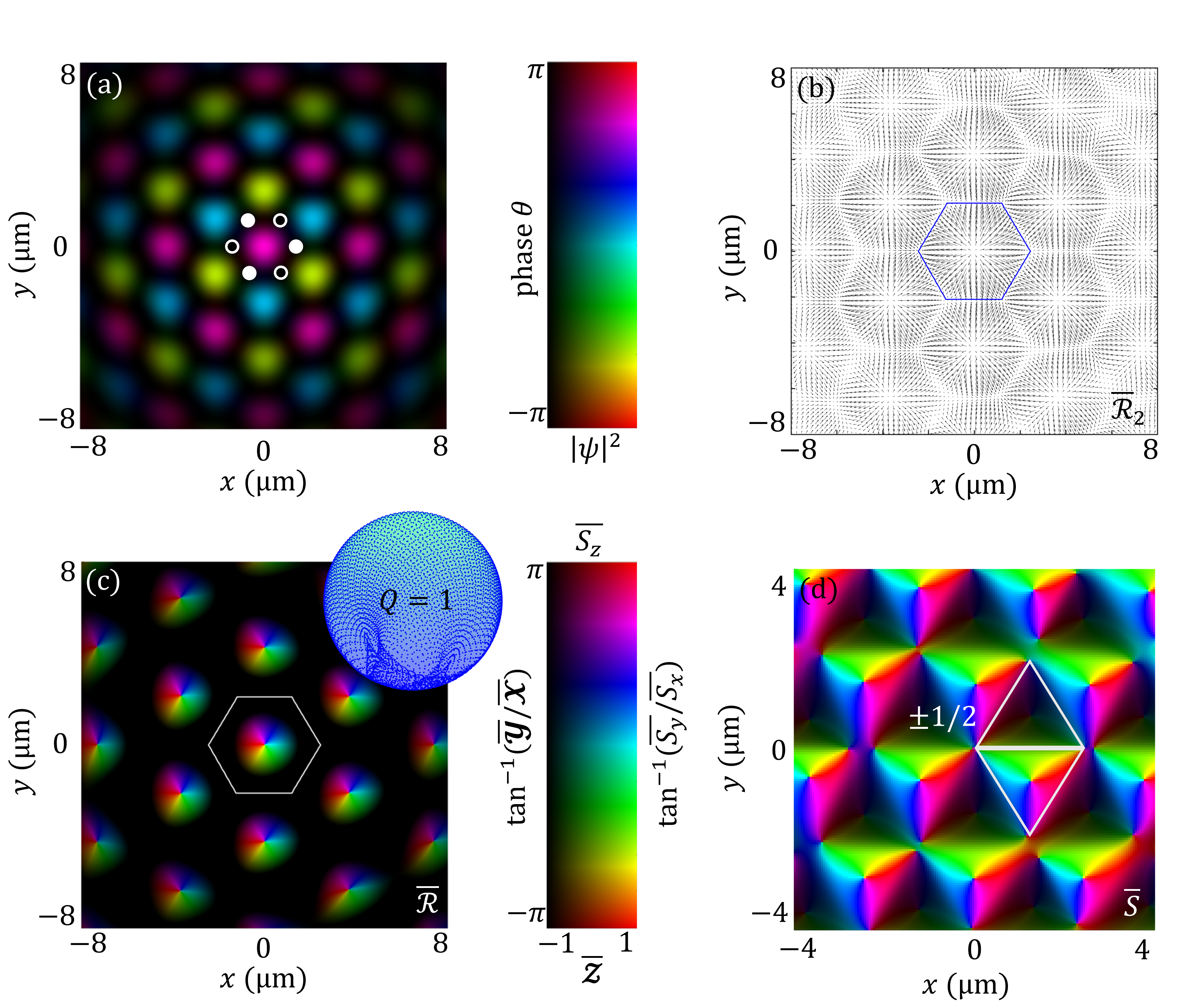}
    \caption{\textbf{Skyrmion pattern under three-wave interference (resonant).} (a) Density (brightness) and phase (color) of the interference pattern. Vortex and anti-vortex pairs are highlighted by circles and dots. (b) In-plane displacement field, with Skyrmion cell marked in blue. (c) Full displacement field, with Skyrmion cell marked in white and unit-sphere mapping shown in the inset together with the Skyrmion number $Q$. (d) Spin density illustrating meron structure.}
    \label{fig:s1}
\end{figure}

\begin{figure}[tb]
    \includegraphics[width=0.75 \textwidth]{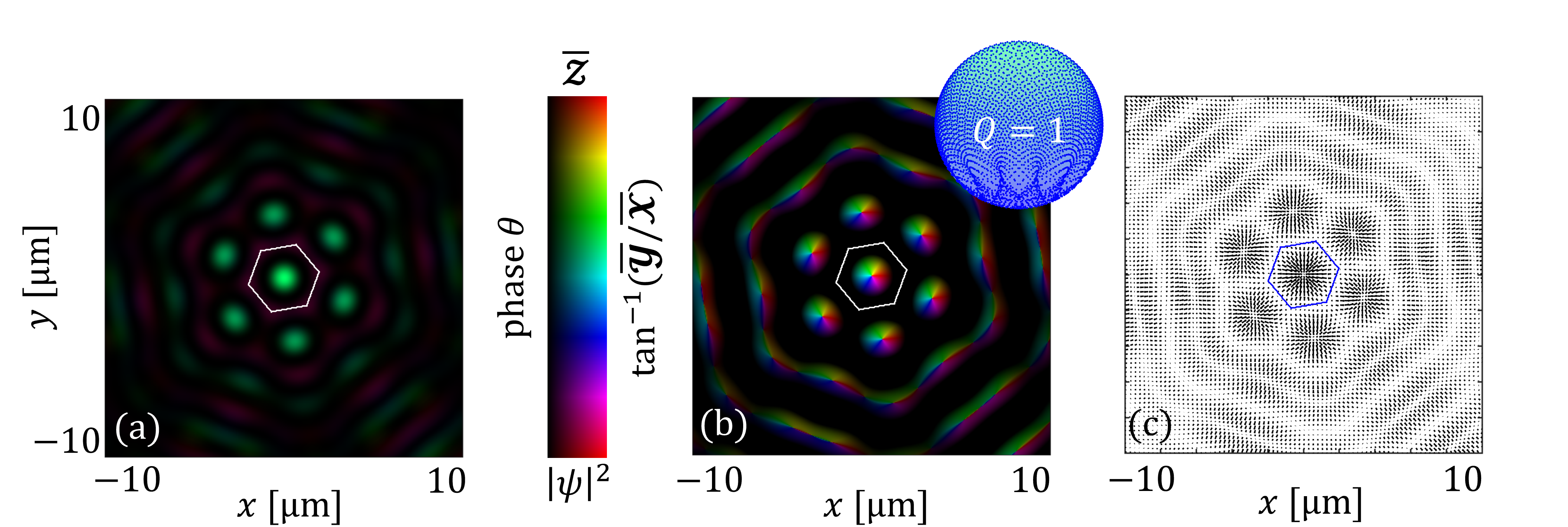}
    \caption{\textbf{Skyrmion bag in moiré lattice (resonant).} (a) Density (brightness) and phase (color) of the interference pattern. (b) Full displacement field, with Skyrmion cell marked in white and unit-sphere mapping shown in the inset together with the Skyrmion number $Q$. (c) In-plane displacement field, with Skyrmion cell marked in blue.}
    \label{fig:s3}
\end{figure}

\begin{figure}[tb]
    \includegraphics[width=0.5 \textwidth]{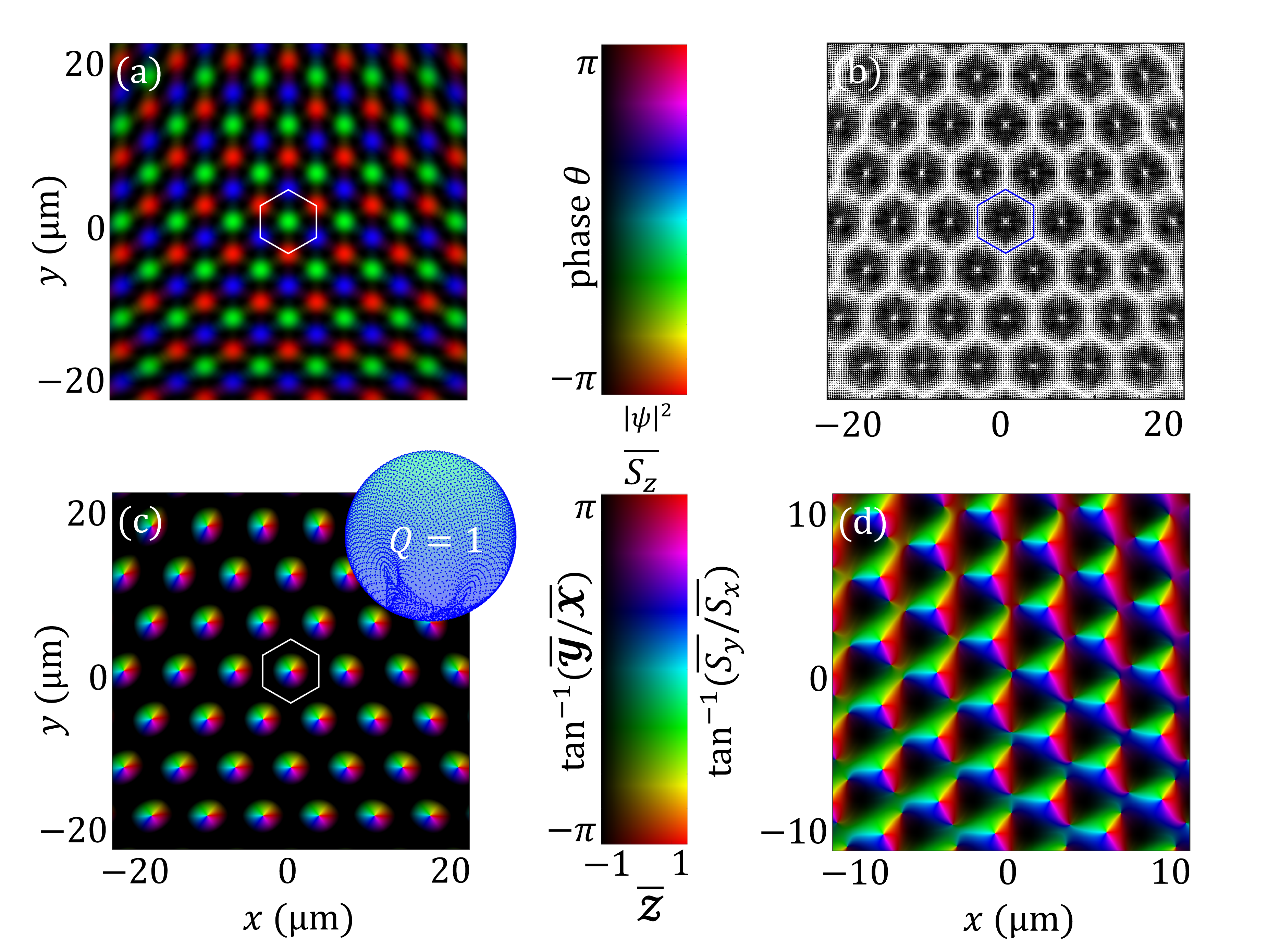}
    \caption{\textbf{Skyrmion pattern under three-wave interference (nonresonant).} (a) Density (brightness) and phase (color) of the interference pattern. (b) In-plane displacement field, with Skyrmion cell marked in blue. (c) Full displacement field, with Skyrmion cell marked in white and unit-sphere mapping shown in the inset together with the Skyrmion number $Q$. (d) Spin density illustrating meron structure.}
    \label{fig:s2}
\end{figure}

\begin{figure}[tbh]
    \includegraphics[width=0.75 \textwidth]{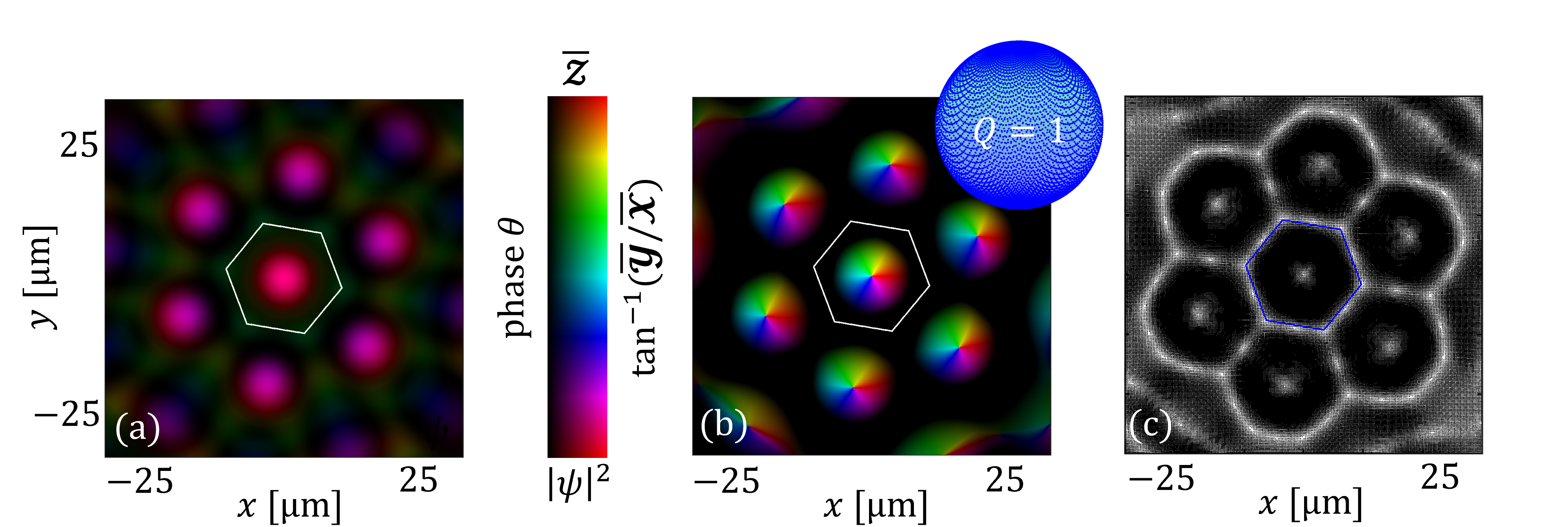}
    \caption{\textbf{Skyrmion bag in moiré lattice (nonresonant).} (a) Density (brightness) and phase (color) of the interference pattern. (b) Full displacement field, with Skyrmion cell marked in white and unit-sphere mapping shown in the inset together with the Skyrmion number $Q$. (c) In-plane displacement field, with Skyrmion cell marked in blue.}
    \label{fig:s4}
\end{figure}

Here we show additional details about the arising Skyrmion patterns in the two excitation scenarios discussed in the main text. Around the system center both resonant and nonresonant excitations lead to quantitatively identical results. Note, that as stated in the main part, nonresonant excitation allows for a more homogeneous density distribution, while the resonantly-driven interference patterns fade faster to the outside. In Fig.~\ref{fig:s1} and \ref{fig:s3} the results for three-wave interference and moiré lattice formation under resonant excitation are shown. In Fig.~\ref{fig:s2} and \ref{fig:s4} the results for three-wave interference and moiré lattice formation under nonresonant excitation are shown. The panels show the wavefunction information, the in-plane displacement field, the full displacement field and (in case of the three-wave interference) the spin density illustrating potential meron structures. These results are in very good agreement with the features measured in water and plasmonic systems as discussed in the main part of our work. 

\newpage
\section{Optical forces in Skyrmion moiré lattice}
\begin{figure}[tb]
    \includegraphics[width=0.6 \textwidth]{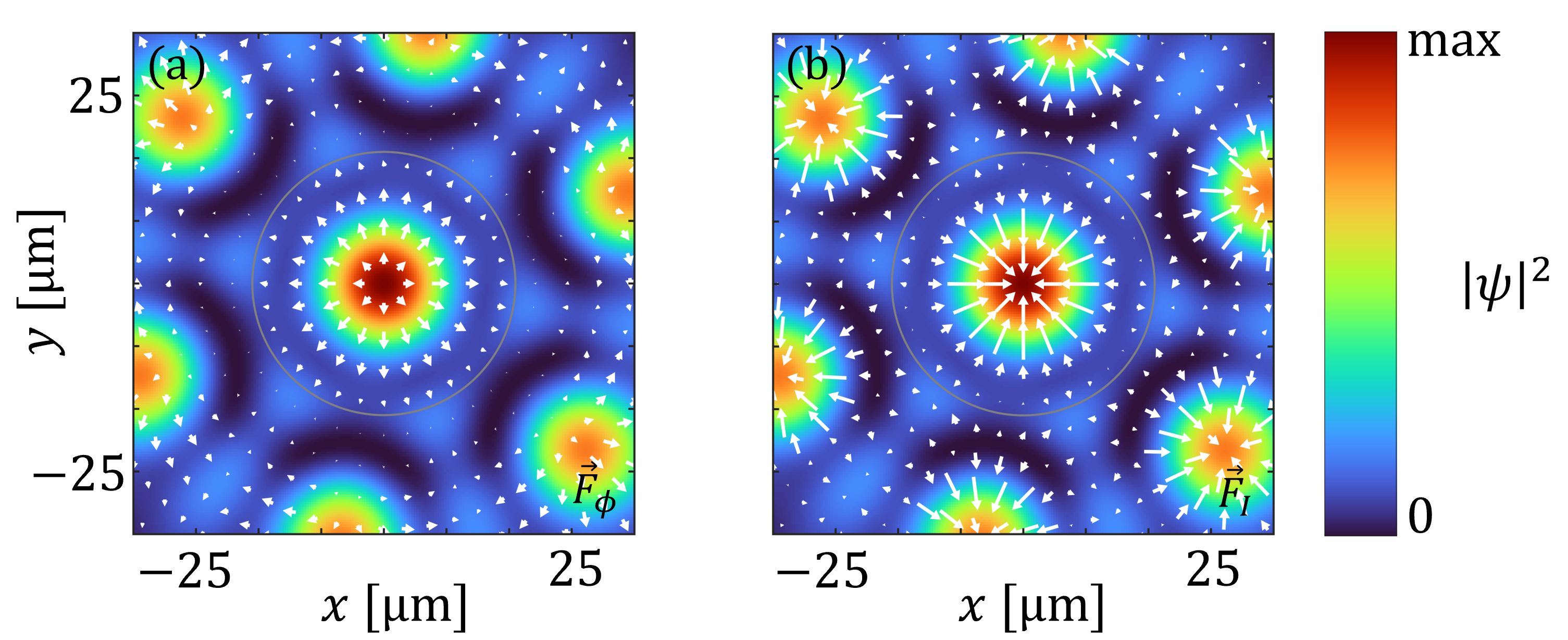}
    \caption{\textbf{Optical forces in Skyrmion moiré lattice.} (a) Phase-gradient force and (b) Density-gradient force as vector fields in front of the density distribution (color) forming the Skyrmion moiré lattice. For visualization the inner bag is displayed.}
    \label{fig:s5}
\end{figure}

In this section the results for the optical forces for non-isolated Skyrmions is introduced. In Fig.~\ref{fig:s5} the optical forces on the Skyrmion moiré lattice are displayed. Notably the density-gradient force $F_I$ forms traps in the Skyrmion center, while the phase-gradient force $F_\phi$ permits trapping between Skyrmions. 

\section{Simplified theoretical model for single Skyrmion case}

Here we present a simplified theoretical description of an isolated Skyrmion under nonresonant pumping, with the aim of enlightening the fundamental physical mechanisms underlying the solution. The starting equations are those in eq. (1) of the main paper. For an incoherent pump $P$ we may set $E=0$. Furthermore, we focus on threshold conditions, where $|\psi|^2 \rightarrow 0$ can be assumed while retaining a finite field amplitude $\psi$. For a cylindrically symmetric pump $P(r)$ we further seek an eigenstate for which 
$$
\psi(r,t) = e^{-i\mu t/\hbar} U(r),
$$
with chemical potential $\mu$. Under steady-state conditions, and neglecting nonlinear effects, the reservoir density is obtained as
$$
n(r) = {P(r)\over\gamma_r}. 
$$
The equation for the mode $U(r)$ then becomes
\begin{equation}\label{Eq2}
\mu U= \left ( -{\hbar^2\over 2 m} \nabla^2 + {g_r\over\gamma_r} P(r)  + {i\hbar\over 2} \left [{R_c\over\gamma_r} P(r)-\gamma_c \right ] \right)  U. 
\end{equation}
This non-Hermitian Schr\"odinger equation (NHSE) can be used to determine the threshold mode $U(r)$. The underlying motivation is that the threshold solution may already reveal key features of the system that persist above threshold, where nonlinear interactions become significant. Skyrmionic structures are known to exist even in passive externally driven cavities. The objective here is therefore to examine the corresponding incoherently pumped system at threshold, before nonlinear effects dominate the dynamics. In this capacity we note that the chemical potential $\mu$ introduced above is complex to allow for loss and gain in the system. At threshold, the imaginary part of the complex chemical potential, $\mu=\mu'+i\mu''$, must vanish, ensuring an exact balance between gain and loss. Otherwise, the mode would either grow exponentially or decay in time.

To derive an approximate analytical solution of the NHSE, we consider a Gaussian pump profile
$$
P(r) = P_0 e^{-r^2/r_p^2} \approx P_0 \left (1-{r^2\over r_p^2} + \ldots \right ) ,
$$
where $P_0$ denotes the peak pump intensity and $r_p$ is the characteristic pump radius. A Taylor expansion around $r=0$ is employed to describe the mode behavior near the center of the pump spot. Using the Taylor expansion eq. (\ref{Eq2}) can be approximated as
\begin{equation}\label{Eq3}
\mu U\approx \left ( -{\hbar^2\over 2 m} \nabla^2 + {g_rP_0 \over\gamma_r}\left (1-{r^2\over r_p^2} \right )+ {i\hbar\over 2} \left [{R_cP_0 \over\gamma_r} \left (1-{r^2\over r_p^2} \right ) -\gamma_c \right ] \right)  U. 
\end{equation}
A key observation is that the resulting equation can be mapped onto a quantum harmonic oscillator with a complex frequency, defined by
\begin{equation}
{1\over 2}m\omega_c^2 = -{g_rP_0\over \gamma_r r_p^2} - {i\hbar\over 2} {R_c P_0\over \gamma_r r_p^2} .
\end{equation}
Then along with the definition
\begin{equation}
\mu^{(0)} =  {g_rP_0\over\gamma_r} + {i\hbar\over 2} \left [{R_cP_0 \over\gamma_r}-\gamma_c \right ],
\end{equation}
the NHSE can be written as
\begin{equation}\label{Eq4}
\mu U= \mu^{(0)} U \underbrace{ -{\hbar^2\over 2 m}\left ( {d^2\over dr^2} +{1\over r} {d\over dr} \right )U  + {1\over 2} m\omega_c^2 r^2 U }.
\end{equation}
The underbraced term in this NHSE is the complex quantum harmonic oscillator with unnormalized Gaussian mode solution
\begin{equation}
U(r) = e^{-r^2/2r_c^2}, \quad r_c = \sqrt{{\hbar\over m\omega_c}} ,
\end{equation}
and complex eigenvalue $\hbar\omega_c$. In this context,
$$
{1\over r_c^2} = {m\omega_c\over \hbar} = {m\omega_c'\over \hbar} + i {m\omega_c''\over \hbar} ,
$$
must be chosen with the root such that $\omega_c' >0$ so that the Gaussian mode
\begin{eqnarray}\label{U}
U(r) &=& e^{-r^2/2r_c^2}\nonumber \\
&=& e^{-{m\omega_c'\over 2\hbar} r^2 - i{m\omega_c''\over 2\hbar} r^2},
\end{eqnarray}
corresponds to a spatially confined mode. With this in mind the complex chemical potential for the model is given by
$$
\mu = \mu^{(0)} + \hbar\omega_c ,
$$
and the threshold condition $\mu''=0$ can be satisfied for a chosen set of parameters by suitable choice of $\gamma_c$.

The central feature of the problem is its non-Hermitian character and the resulting modification of the mode structure. For the usual real quantum harmonic oscillator the modes can be chosen to have uniform phases, but here this is not possible even for the Gaussian mode, this is akin to the modes of so called gain guided semiconductor lasers where the oscillating mode has a phase curvature. The phase curvature associated with $\omega_c''\neq 0$ plays a crucial role in enabling the Skyrmionic structure discussed below.

Next, we analyze the components of the real-valued displacement field and relate them to the behavior shown in Fig.~2 of the main manuscript for the Gaussian solution derived above. First, the density profile in Fig. 2(a) will have the radial variation $\propto e^{-{m\omega_c'\over \hbar} r^2}$. The components of the real valued displacement field (for $t=0$) are:

\begin{itemize}

\item the x-component
\begin{eqnarray}\label{X}
{\cal X} &\propto&  \Re \left ( {\partial U\over \partial x}  \right ) \propto x e^{-{m\omega_c'\over 2\hbar} r^2} \left [ \cos\left ( {m\omega_c''\over 2\hbar} r^2 \right ) + {\omega_c''\over \omega_c'} \sin\left ( {m\omega_c''\over 2\hbar} r^2 \right ) \right ]  \nonumber \\
&\propto& x e^{-{m\omega_c'\over 2\hbar} r^2} \cos\left ( {m\omega_c''\over 2\hbar} r^2 -\alpha \right ) ,
\end{eqnarray}
where $\alpha = \tan^{-1}  {\omega_c''\over \omega_c'} $ ,

\item the y-component

\begin{eqnarray}\label{Y}
{\cal Y} &\propto&  \Re \left ( {\partial U\over \partial y}  \right ) \propto y e^{-{m\omega_c'\over 2\hbar} r^2} \left [ \cos\left ( {m\omega_c''\over 2\hbar} r^2 \right ) + {\omega_c''\over \omega_c'} \sin\left ( {m\omega_c''\over 2\hbar} r^2 \right ) \right ]  \nonumber \\
&\propto& y e^{-{m\omega_c'\over 2\hbar} r^2} \cos\left ( {m\omega_c''\over 2\hbar} r^2 -\alpha \right ),
\end{eqnarray}
\item and the z-component

\begin{equation}\label{Z}
{\cal Z} \propto \Re  (U) \propto e^{-{m\omega_c'\over 2\hbar} r^2} \cos\left ( {m\omega_c''\over 2\hbar} r^2 \right ).
\end{equation}

\end{itemize}
In each component, the trigonometric terms account for the radial sign reversals observed in the field distribution, as illustrated for ${\cal Z}$ in Fig.~2(c) of the main text. Similarly, the prefactors appearing in ${\cal X}$ and ${\cal Y}$ have the form of first-order Hermite-Gaussian modes along the $x$ and $y$ directions. This explains why ${\cal R}_2$ in Fig.~2(d) vanishes at the origin, while the cosine modulation accounts for the additional nodal ring structure. The Gaussian approximation should not be interpreted quantitatively for radii significantly larger than the pump radius $r_p$. This limitation arises from the parabolic approximation of the gain profile, which introduces nonphysical absorption at sufficiently large distances from the pump center.

The purpose of the simplified model is to demonstrate that the Skyrmionic character of the solutions is intimately connected to the distinctive modal properties of the NHSE. In particular, the analysis links the emergence of Skyrmionic features to the phase curvature of the non-Hermitian eigenmodes, a quantity that is, in principle, experimentally accessible. The model is not intended to provide quantitatively precise predictions, rather, its value lies in the physical insight it offers and in its potential to reveal how the modal properties scale with the relevant system parameters. Although the approach could be extended through a variational Gaussian treatment, such a refinement is unlikely to yield substantially deeper physical insight.

\section{Skyrmion number as a function of Skyrmion cell radius}
\begin{figure}[tb]
    \centering
    \includegraphics[width=0.4\linewidth]{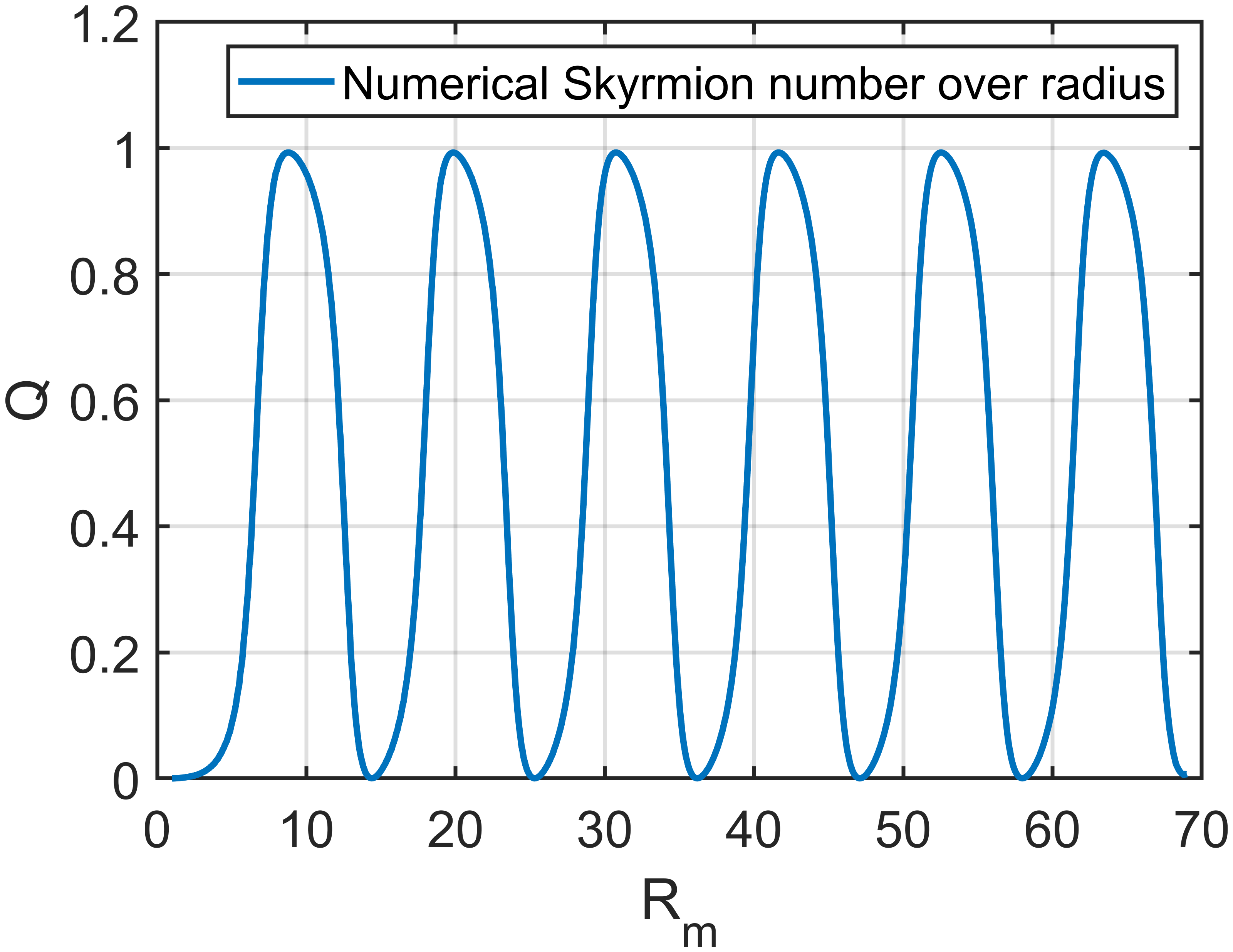}
    \caption{Calculated Skyrmion number $Q$ as a function of the Skyrmion cell radius $R_m$. Due to the normalization of the displacement field, the resulting topological charge oscillates between $Q=0$ and $Q=1$ depending on the selected integration radius.}
    \label{fig:QvsR}
\end{figure}

Here we discuss the scaling of the Skyrmion number with the radius of the circular integration domain in case of the single spontaneous Skyrmion formed in Fig. 2 of the main manuscript. To visualize the dependence of the Skyrmion number on the chosen integration domain size, Figure \ref{fig:QvsR} shows the cell radius as a function of the resulting Skyrmion number. The Skyrmion number was calculated using the normalized displacement field. The computed value oscillates between $Q = 0$ and $Q = 1$ depending on the selected radius.

Importantly, the radius used for the Skyrmion number calculation is not chosen arbitrarily, nor is it adjusted iteratively until $Q = 1$ is obtained. Instead, the radius is determined from the first minimum of the displacement field component $\mathcal{Z}$. In practice, the extrema of the displacement field are used to define the boundaries of the Skyrmion cell, ensuring that the smallest physically meaningful cell is selected for the calculation.

\end{document}